\documentclass[%
reprint,
superscriptaddress,
amsmath,amssymb,
aps,
]{revtex4-2}

\usepackage{graphicx}
\usepackage{dcolumn}
\usepackage{bm}

\begin{document}
\title{Anomalously field-susceptible spin soft-matter emerging in an electric-dipole liquid candidate}
\author{Mizuki Urai}
\email{urai@mdf2.t.u-tokyo.ac.jp}
\affiliation{Department of Applied Physics, University of Tokyo, Tokyo 113-8656, Japan.}
\author{Kazuya Miyagawa}
\affiliation{Department of Applied Physics, University of Tokyo, Tokyo 113-8656, Japan.}
\author{Yuta Watanabe}
\affiliation{Department of Applied Physics, University of Tokyo, Tokyo 113-8656, Japan.}
\author{Elena I. Zhilyaeva}
\affiliation{Institute of Problems of Chemical Physics RAS, 142432 Chernogolovka, Russia.}
\author{Svetlana A. Torunova}
\affiliation{Institute of Problems of Chemical Physics RAS, 142432 Chernogolovka, Russia.}
\author{Rimma N. Lyubovskaya}
\altaffiliation{Deceased.}
\affiliation{Institute of Problems of Chemical Physics RAS, 142432 Chernogolovka, Russia.}
\author{Natalia Drichko}
\affiliation{The Institute for Quantum Matter and the Department of Physics and Astronomy, The Johns Hopkins University, Baltimore, MD 21218, USA.}
\affiliation{The Institute for Solid State Physics, The University of Tokyo, Kashiwa, Chiba 277-8581, Japan.}
\author{Kazushi Kanoda}
\email{kanoda@ap.t.u-tokyo.ac.jp}
\affiliation{Department of Applied Physics, University of Tokyo, Tokyo 113-8656, Japan.}

\date{\today}

\begin{abstract}
Mutual interactions in many-body systems bring about a variety of exotic phases, among which liquid-like states failing to order due to frustration are of keen interest. 
Recently, an organic system with an anisotropic triangular lattice of molecular dimers has been suggested to host a dipole liquid arising from intradimer charge-imbalance instability, possibly offering an unprecedented stage for the spin degrees of freedom. 
Here we show that an extraordinary unordered(unfrozen) spin state having soft-matter-like spatiotemporal characteristics is substantiated in this system. 
$^{1}$H NMR spectra and magnetization measurements indicate that gigantic, staggered moments are non-linearly and inhomogeneously induced by magnetic field whereas the moments vanish in the zero-field limit. 
The analysis of the NMR relaxation rate signifies that the moments fluctuate at a characteristic frequency slowing down to below MHz at low temperatures. 
The inhomogeneity, local correlation, and slow dynamics indicative of middle-scale dynamical correlation length suggest a novel frustration-driven spin clusterization. 
\end{abstract}

\maketitle

\section{Introduction}
\begin{figure*}
	\includegraphics{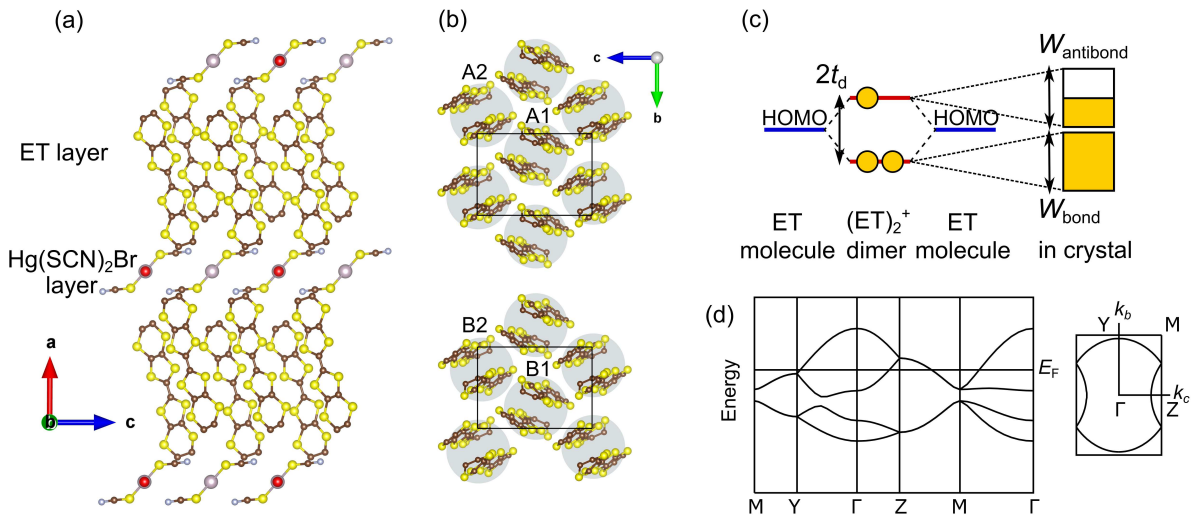}
	\caption{\label{fig:structure} (a) The crystal structure of $\kappa$-Hg-Br~\cite{BRAS-1992-Konovalikhin}, crystalizing in $C2/c$ space group. The ET and Hg(SCN)$_{2}$Br layers alternately stack in the crystal. There are two kinds of ET layers (A and B). The top views of the conducting ET layers are shown in (b). Each pair of the ET molecules facing each other (highlighted in grey) forms a dimer. The unit cell contains four dimers:The two dimers A1(B1) and A2(B2) in a layer are mutually related by the glide operation with the glide $ac$ plane. The two dimers A1(A2) and B1(B2) are mutually related by the translational operator across a Hg(SCN)$_{2}$Br layer. The solid lines in (b) represent a unit cell. Panels (a) and (b) were drawn using the visualizing software VESTA~\cite{VESTA}. (c) Schematic electronic structure of $\kappa$-Hg-Br. In $\kappa$-(ET)$_{2}$\textit{X}, two ET highest-occupied molecular orbitals (HOMOs) in a dimer form bonding and antibonding orbitals through the intradimer transfer energy $t_{d}$, and each of them forms a conduction sub-band with the bandwidth, $W_{\textrm{bond}}$ or $W_{\textrm{antibond}}$. In the case of $\kappa$-Hg-Br, the band gap between the bonding and antibonding bands is narrow or vanishing, as seen in (d). (d) The band structure (left) and Fermi surface (right) of $\kappa$-Hg-Br based on the reported crystal structure~\cite{BRAS-1992-Konovalikhin}. The range of vertical axis in the left panel is from –0.7 to 0.7 eV. The transfer integrals and band dispersion were obtained by the extended H\"{u}ckel method and tight-binding calculation, using the program package~\cite{BullChem-1984-Mori}.}
\end{figure*}

Strongly correlated electron systems harbor a rich variety of unconventional phenomena with interactions and quantum nature strongly entangled~\cite{RMP-1998-Imada}. 
Among them is the quantum spin liquid (QSL), in which spins are highly correlated with one another but do not order due to strong zero-point fluctuations. A quest for QSLs has been a focus of profound interest in the physics of interacting quantum many body systems~\cite{RMP-2017-Zhou,RPP-2017-Savary,Science-2020-Broholm,AnnRev-2019-Knolle}. 
The layered organic conductors, $\kappa$-(ET)$_{2}$\textit{X}, where ET denotes bis(ethylenedithio)tetrathiafulvalene and \textit{X} is monovalent anion species, have been studied as model systems exhibiting the Mott metal-insulator transition and including the QSL candidates~\cite{AnnRev-2011-Kanoda,ChemRev-2004-Miyagawa}. 
A key ingredient for understanding the electronic states in $\kappa$-(ET)$_{2}$\textit{X} is the strongly dimerized ET molecules~\cite{JPSJ-1996-Kino}; the ET dimers are arranged to form anisotropic triangular lattices (Fig.~\ref{fig:structure}). 
Given that the ET dimer is regarded as a supermolecule occupying a lattice site, the complicated crystal structure is modeled by an anisotropic triangular lattice hosting one hole with a spin-1/2 on each lattice site. 
This simplification is a basis for understanding magnetism in the Mott insulator state of $\kappa$-(ET)$_{2}$\textit{X}, including antiferromagnets $\kappa$-(ET)$_{2}$Cu[N(CN)$_{2}$]Cl~\cite{PRL-1995-Miyagawa,PRB-2003-Smith,JPSJ-2018-Ishikawa,PRB-2008-Kagawa} and deuterated $\kappa$-(ET)2Cu[N(CN)2]Br~\cite{PhysicaB-2000-Miyagawa,PRB-2020-Oinuma}, and possible QSLs $\kappa$-(ET)$_{2}$Cu$_{2}$(CN)$_{3}$~\cite{PRL-2003-Shimizu} and $\kappa$-(ET)$_{2}$Ag$_{2}$(CN)$_{3}$~\cite{PRL-2016-Shimizu} (abbreviated as $\kappa$-Cu-Cl, $\kappa$-Cu-Br, $\kappa$-Cu-CN, and $\kappa$-Ag-CN hereafter, respectively).
The discussed above simple modeling of the lattice neglects the internal degrees of freedom within the ET dimers; however, recent theoretical calculations have proposed that the intradimer electronic degrees of freedom can play a vital role in the emergence of exotic states in the nominal ``dimer Mott insulators''~\cite{PRB-2010-Hotta,PRB-2016-Naka}. 
For instance, the relaxer-type dielectric anomaly observed in $\kappa$-Cu-CN~\cite{PRB-2010-Abdel} is argued to be a manifestation of the intradimer charge imbalance in the QSL state under debate~\cite{PRB-2010-Abdel,JPSJ-2015-Yakushi,PRL-2013-Itoh,PRR-2020-Kobayashi}.
More recently, $\kappa$-(ET)$_{2}$Hg(SCN)$_{2}$Br ($\kappa$-Hg-Br) [Fig.~\ref{fig:structure}(a) and(b)] has been suggested to host a quantum dipole liquid where electric dipoles presumably created by intradimer charge imbalance keep fluctuating down to low temperatures~\cite{Science-2018-Hassan}. 
The $\kappa$-Hg-Br undergoes a first-order metal-insulator transition around $T_{\textrm{MI}} \sim$80--90~\cite{PRB-2017-Ivek} without a clear evidence of charge order below $T_{\textrm{MI}}$ in the optical and dielectric measurements~\cite{Science-2018-Hassan,PRB-2017-Ivek} in contrast to an isostructural sister compound $\kappa$-(ET)$_{2}$Hg(SCN)$_{2}$Cl ($\kappa$-Hg-Cl), which undergoes a charge-ordering metal-insulator transition at $T_{\textrm{CO}} \sim$30 K~\cite{PRB-2014-Drichko,PRL-2018-Gati}. 
Therefore, $\kappa$-Hg-Br is possibly at the phase border between Mott insulator and charge ordering. 
Indeed, the ET dimerization in $\kappa$-Hg-Br is weak among $\kappa$-(ET)$_{2}$\textit{X} so that the band profile has a dual character, a quarter-filled band of ET molecular orbitals and a half-filled band of the dimer (ET$_{2}$) antibonding orbitals [Fig.~\ref{fig:structure}(c)], as suggested by the extended H\"{u}ckel tight-binding band calculations predicting a narrow or vanishing band gap [Fig.~\ref{fig:structure}(d)]. 
The former favors a charge ordered insulator whereas the latter favors a Mott insulator. 
The spin state in such a marginal situation can be nontrivial because the spin-exchange interactions are affected by the charge fluctuations; interestingly, the time scale of charge fluctuations in $\kappa$-Hg-Br suggested by the motional narrowing analysis of the Raman spectra is ~1 THz, which is of the same order of the typical spin exchange frequency in $\kappa$-type ET compounds~\cite{Science-2018-Hassan}. 
So far, the magnetization has been reported to show nonlinear soft ferromagnetic-like behavior~\cite{PRB-2018-Hemmida} with no hysteresis~\cite{npj-2021-Yamashita}. 
The magnetization, \textit{e.g.}, at a magnetic field of 1 T, is about one order of magnitude larger than that of the well-studied typical dimer Mott insulators $\kappa$-Cu-Cl~\cite{PRB-2003-Smith,JPSJ-2018-Ishikawa} and $\kappa$-Cu-Br~\cite{PRB-2020-Oinuma}. 
More recently, a $^{13}$C NMR study~\cite{PRB-2020-Le} has revealed that the spin state in $\kappa$-Hg-Br is highly disordered. 
The microscopic spin state to give such extraordinary magnetic properties has yet to be clarified.
The present work utilizes $^{1}$H NMR spectroscopy with the magnetic field varied over a 20-fold range from 0.30 to 6.00 T, combined with spectral simulations, to make clear the microscopic local moment configuration and its field and temperature evolution in $\kappa$-Hg-Br. 
$^{1}$H NMR is superior in capturing the widespread spectra at high fields to $^{13}$C NMR which has difficulty in acquiring the entire spectrum owing to the large hyperfine coupling. 
Here, we report that an anomalously field-susceptible unordered(unfrozen) spin state with highly correlated characteristics emerges in $\kappa$-Hg-Br.

\section{Experimental Methods}
The samples of $\kappa$-Hg-Br were synthesized by the electrochemical method. 
In the $^{1}$H NMR experiments, we applied magnetic fields nearly parallel to the $a$ axis of a single crystal weighing 0.31 mg and measured the NMR spectra and nuclear spin-lattice relaxation time, $T_{1}$, which probe the near-static and dynamical spin states, respectively. 
The NMR spectra were acquired by the standard solid-echo method and the nuclear spin-lattice relaxation time, $T_{1}$, was determined by the standard saturation-recovery method.
We measured the magnetization for an assembly of four single crystals weighing 2.06 mg in total, using a Magnetic Property Measurement System (MPMS-5S, Quantum Design Inc.). 
The first-order metal-insulator transition was signaled by a jump in magnetization at the $T_{\textrm{MI}}$ of 90 K. 
In all the data shown here, the core diamagnetic contribution of $-5.15\times10^{-4}$ emu/mol was subtracted.

\section{Results}
In this section, we first show the $^{1}$H NMR spectra and compare the magnitude of local fields with the magnetization. After that, we discuss the spin dynamic along with the results of $^{1}$H nuclear spin-lattice relaxation.
\subsection{$^{1}$H NMR spectra}
\begin{figure}
	\includegraphics{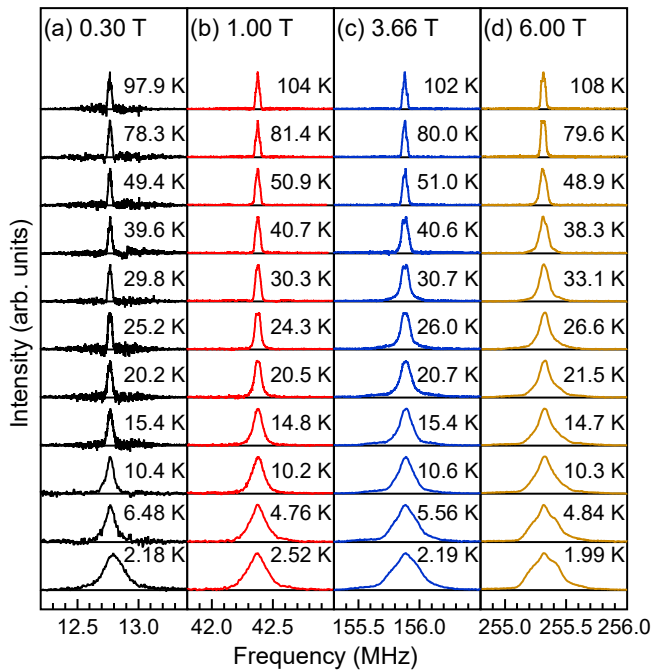}
	\caption{\label{fig:spectra} $^{1}$H NMR spectra measured in the magnetic fields of 0.30, 1.00, 3.66 and 6.00 T.}
\end{figure}

Figures~\ref{fig:spectra}(a) display the $^{1}$H NMR spectra of $\kappa$-Hg-Br in the magnetic fields of 0.30, 1.00, 3.66 and 6.00 T. 
At high temperatures above $T_{\textrm{MI}}$, 90 K, the temperature-insensitive spectral shape is determined by the $^{1}$H nuclear dipolar interactions. 
Upon cooling, the spectral lines are symmetrically broadened in every field. 
A structureless broadened line shape indicates an inhomogeneous distribution of the local fields in contrast to discrete spectral lines as observed in the commensurate antiferromagnet, $\kappa$-Cu-Cl~\cite{PRL-1995-Miyagawa}. 
Incommensurate magnetic order is ruled out, as it would give the distinctive lineshape with wings on the edges. 
At the lowest measured temperature of $\sim$2 K, the spectra are extended to $\pm$200 kHz, which exceeds the spectral width, $\pm$100 kHz, observed in $\kappa$-Cu-Cl with an antiferromagnetic moment of 0.45$\mu_{\textrm{B}}$/dimer~\cite{ChemRev-2004-Miyagawa} with the Bohr magneton $\mu_{\textrm{B}}$, suggesting that the maximum moment amounts to $\sim$1$\mu_{\textrm{B}}$/dimer or larger, as discussed in details below.
We characterize the distribution of local fields by the square root of the second moment of the spectra, $\langle \Delta f^{2}\rangle^{1/2}$, which is a measure of the spread of the local fields projected along the field direction. 
The $\langle \Delta f^{2}\rangle$ is approximated by a sum of the temperature ($T$) and field ($H$) insensitive nuclear dipolar contribution $\langle \Delta f^{2}\rangle_{\textrm{n}}$ and the electron moment contribution $\langle \Delta f^{2}\rangle_{\textrm{m}}$. 
Above $T_{\textrm{MI}}$, the NMR spectra are solely determined by the nuclear dipolar interaction, so, we use the $\langle \Delta f^{2}\rangle$ value at 100 K ($> T_{\textrm{MI}}$) as $\langle \Delta f^{2}\rangle_{\textrm{n}}$. 

Figure~\ref{fig:spectra}(b) displays the temperature profiles of $\langle \Delta f^{2}\rangle_{\textrm{m}}^{1/2}$ given by $(\langle \Delta f^{2}\rangle - \langle \Delta f^{2}\rangle_{\textrm{n}})^{1/2}$, which prominently increases below 15, 25, 30--40, and 40--50 K in the fields of 0.30, 1.00, 3.66, and 6.00 T, respectively. 
\begin{figure}
	\includegraphics{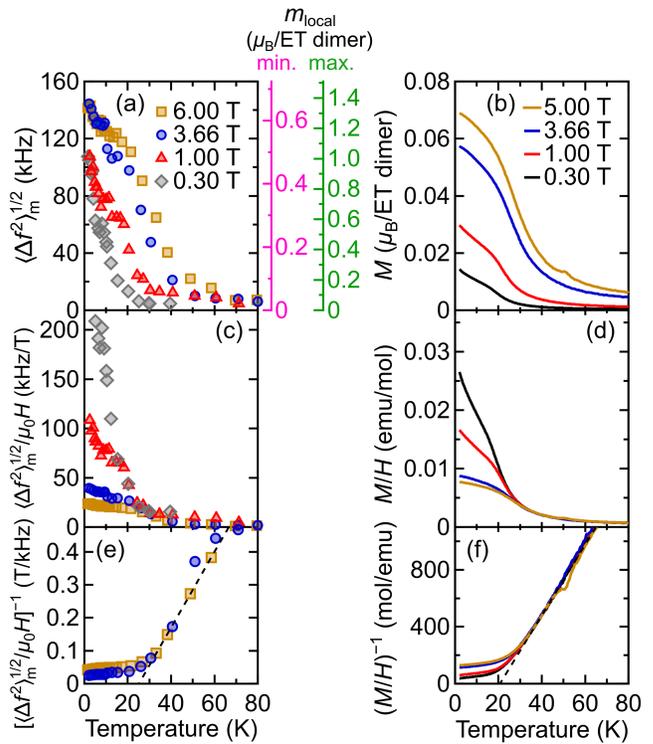}
	\caption{\label{fig:MT_curves} (a,b) Temperature dependence of (a) the square root of the electron moment contribution to second moment of NMR spectra, $\langle \Delta f^{2}\rangle_{\textrm{m}}^{1/2}$, and (b) magnetization, $M$. The right axes in (a) show the scales for the minimum (min.) and maximum (max.) cases of the local moment, $m_{\textrm{local}}$, estimated in the present work (see the main text for the estimation in details). (c,d) Temperature profiles of (c) $\langle \Delta f^{2}\rangle_{\textrm{m}}^{1/2}/\mu_{0}H$ and (d) $M/H$, and their inverse [(e) and (f)].}
\end{figure}
The temperature variation of $\langle \Delta f^{2}\rangle_{\textrm{m}}^{1/2}$ is similar to that of the magnetization, $M$ [Fig.~\ref{fig:MT_curves}(b)], indicating that local moments develop along with the increase of magnetization. 
Figures~\ref{fig:MT_curves}(c) and (d) display $\langle \Delta f^{2}\rangle_{\textrm{m}}^{1/2}/\mu_{0}H$ and $M/H$, where $\mu_{0}$ denote the magnetic constant, both of which are field-dependent below $\sim$30 K and, above 30 K, vary with the Weiss temperature of 20--25 K (Insets).
To find the configuration and magnitude of the local moments, we simulate the $^{1}$H NMR spectra for several conceivable cases of collinear spin configurations. 
Because the $^{1}$H's in ET has negligibly small Fermi contact interactions, the local fields at the 1H sites are calculated by adding up dipolar fields from electronic moments on the constituent atomic sites. 
Using the Mulliken populations for the spin density distribution in the ET molecule~\cite{BullChem-1984-Mori} and the crystal structure for the ET arrangement~\cite{BRAS-1992-Konovalikhin}, we summed up the dipolar contributions from the atomic sites within a sphere of $\sim$100 {\AA} in radius (see Appendix~\ref{appendix:SPCsimulation} for details). 

\begin{figure*}
	\includegraphics{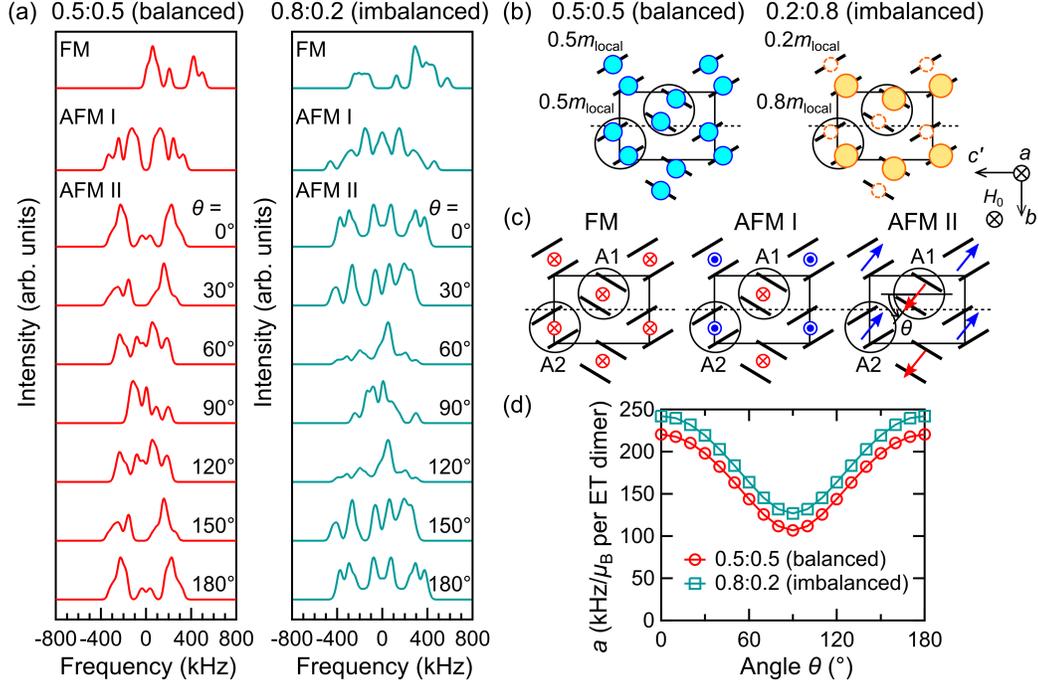}
	\caption{\label{fig:SPCsimulation} (a) The calculated spectra with the moment of 1$\mu_{\textrm{B}}$ on an ET dimer for the ferromagnetic (FM) configuration [left panel of (c)] and the antiferromagnetic (AFM) ones with staggered moments directed parallel (AFM I) and perpendicular (AFM II) to the applied field (normal to the layers) [middle and right panels of (c)]. The left and right panels of (a) are the results for the cases with the intradimer spin-density disproportionation of 0.5:0.5 and 0.8:0.2, depicted in (b), respectively. For AFM II, the moment direction is varied in the perpendicular plane. (d) The angular dependence of the coefficient $a$, which relates the linewidth $\langle \Delta f^{2}\rangle_{\textrm{m}}^{1/2}$ and the size of the moment $m_{\textrm{local}}$ in the form of $\langle \Delta f^{2}\rangle_{\textrm{m}}^{1/2} = am_{\textrm{local}}$ for AFM II. The results shown here assume that AFM I and II has inter-layer ferromagnetic coupling. The values of $\langle \Delta f^{2}\rangle_{\textrm{m}}^{1/2}$ for the cases with inter-layer antiferromagnetic couplings are shown in Appendix~\ref{appendix:SPCsimulation}.}
\end{figure*}

Figure~\ref{fig:SPCsimulation} shows the calculated spectra with the moment of 1$\mu_{\textrm{B}}$ on an ET dimer for the ferromagnetic (FM) configuration and the antiferromagnetic (AFM) configurations with staggered moments directed parallel (AFM I) and perpendicular (AFM II) to the applied field (normal to the layers); for AFM II, the moment direction is varied in the perpendicular plane. 
The FM configuration gives a spectrum shifted to one side, contradicting the observation; this case is unambiguously ruled out. 
Each of AFM I and II has inter-layer ferromagnetic and antiferromagnetic cases. 
Although the spectra in the AFM II cases with the moments on the dimers aligned mutually parallel between the neighboring layers are slightly asymmetric owing to the glide symmetry of the crystal structure, all the simulated spectra in the AFM I and II are roughly symmetrical and are able to explain the observation given an inhomogeneous distribution in the magnitude of the moment. 
Thus, it turns out that Figs.~\ref{fig:MT_curves}(a) and (b) compare the staggered and uniform moments and Figs.~\ref{fig:MT_curves}(c) and (d) compare their susceptibilities.
To evaluate the magnitudes of the moments in $\kappa$-Hg-Br with reference to the simulations, we calculated the coefficient, $a$, relating the linewidth $\langle \Delta f^{2}\rangle_{\textrm{m}}^{1/2}$ and the size of the moment $m_{\textrm{local}}$ in the form of $\langle \Delta f^{2}\rangle_{\textrm{m}}^{1/2} = am_{\textrm{local}}$ for each spin configuration. 
For AFM I, $a = 2.0\times10^{2}$ kHz/$\mu_{\textrm{B}}$ per ET dimer, and for AFM II, $a$ varies in a range of 1.1--2.2$\times10^{2}$ kHz/$\mu_{\textrm{B}}$ per ET dimer depending on the field direction (see Appendix~\ref{appendix:SPCsimulation} for details). 
With this range for $a$, the $\langle \Delta f^{2}\rangle_{\textrm{m}}^{1/2}$ value of 108 kHz at 1.7 K in 0.30 T corresponds to the value of moments, $m_{\textrm{local}}$ = 0.5--1.0$\mu_{\textrm{B}}$ on an ET dimer (or 0.25--0.5$\mu_{\textrm{B}}$ on an ET), which is comparable to or larger than the AFM moments in other $\kappa$-(ET)$_{2}$\textit{X}~\cite{PRL-1995-Miyagawa,PRB-2003-Smith,JPSJ-2018-Ishikawa,PRB-2008-Kagawa,PhysicaB-2000-Miyagawa,PRB-2020-Oinuma}. 
We also considered the hypothetical case of the intradimer spin density imbalance. The simulated spectra with the disproportionation of 0.8:0.2 take different shapes from the 0.5:0.5 case discussed above; however, the two cases are nearly indistinguishable in the second moment and therefore the values of $a$. 
Thus, the emergence of large moments in the applied fields is concluded whether the spin density is imbalanced or not in a dimer.
The spectral shape reflects the distribution profile of the field-induced staggered moments. 

\begin{figure}
	\includegraphics{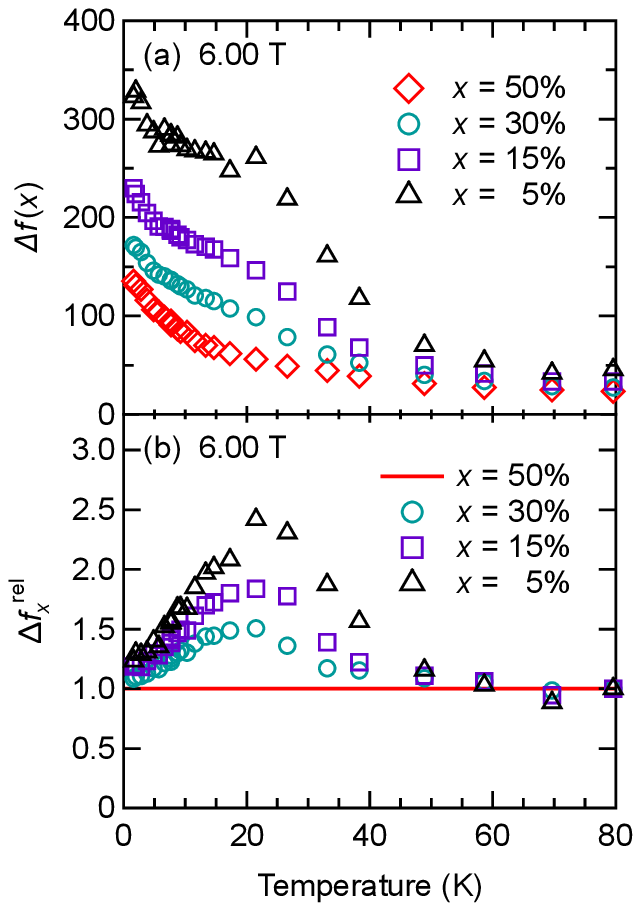}
	\caption{\label{fig:Linewidth} Temperature dependence of the $^{1}$H NMR spectral shape in the field of 6.00 T. The NMR linewidths determined by the half-widths at 50, 30, 15, and 5\% of the maximum value, $\Delta f_{50\%}$, $\Delta f_{30\%}$, $\Delta f_{15\%}$, and $\Delta f_{5\%}$. (Inset) Relative linewidth, $\Delta f^{rel}_{x}$ ($x$ = 50\%, 30\%, 15\%, and 5\%), normalized at 80 K (see the text for the definition).}
\end{figure}

Figure~\ref{fig:Linewidth} shows the temperature dependence of the spectral widths in 6.00 T, defined by the half-widths at 5\%, 15\%, 30\% and 50\% of the maximum value, $\Delta f_{5\textrm{\%}}$, $\Delta f_{15\textrm{\%}}$, $\Delta f_{30\textrm{\%}}$ and $\Delta f_{50\textrm{\%}}$, each of which exhibits distinctive temperature variation.
On cooling, the $\Delta f_{5\%}$ appreciably increases from 50 K and saturates below 20 K, while this feature is less clear in $\Delta f_{15\textrm{\%}}$ and further so in $\Delta f_{30\textrm{\%}}$, and then $\Delta f_{50\textrm{\%}}$ no longer has the convex structure around 20 K but conversely accelerates its increase below 20 K. 
These features are better characterized by the relative widths of $\Delta f_{x}$ ($x$ = 5\%, 15\% and 30\%) to $\Delta f_{50\textrm{\%}}$, $\Delta f^{rel}_{x}(T) \equiv \Delta f_{x}(T)/\Delta f_{50\textrm{\%}}(T)$, plotted in the inset of Fig.~\ref{fig:Linewidth}. 
This illustrates that the spectral tail part, namely, the maximum range of the distributed moments increases upon cooling down to 20 K prior to the growth of the moments on average, which follows below 20 K. 

\begin{figure}
	\includegraphics{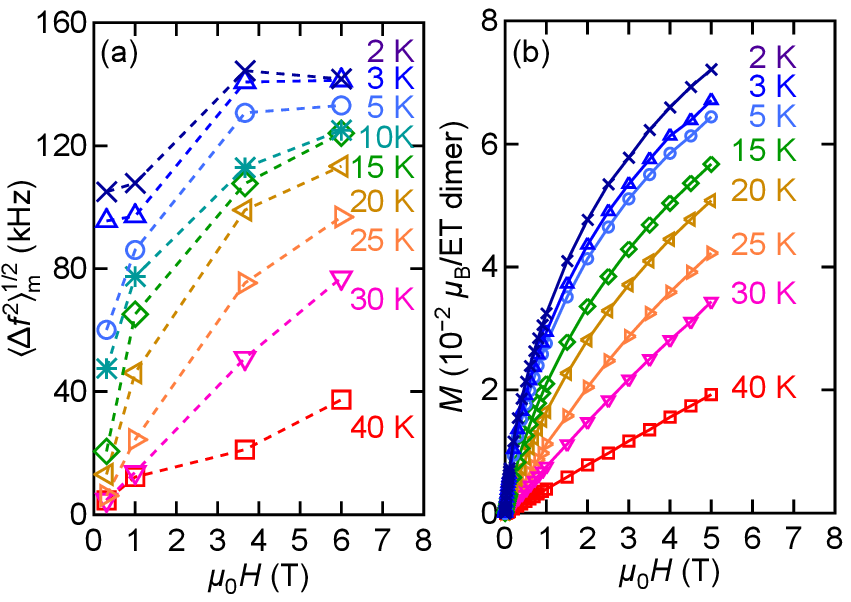}
	\caption{\label{fig:MH_curves} Field dependence of (a) the electron moment contribution to second moment of NMR spectra, $\langle \Delta f^{2}\rangle_{\textrm{m}}^{1/2}$, and (b) magnetization at fixed temperatures.}
\end{figure}

Figure~\ref{fig:MH_curves}(a) shows the field dependence of $\langle \Delta f^{2}\rangle_{\textrm{m}}^{1/2}$. 
At 30 and 40 K, it linearly varies with the applied field, and, at lower temperatures, the variation becomes nonlinear. 
Except at 2 and 3 K, the $\langle \Delta f^{2}\rangle_{\textrm{m}}^{1/2}$ appears to approach zero in the zero-field limit, consistent with the magnetization behavior shown in Fig.~\ref{fig:MH_curves}(b); that is, the magnetic moments nonlinearly growing with magnetic field vanish in the zero-field limit, namely being paramagnetic in nature. 
Remarkably, however, the staggered moments are more than an order of magnitude larger than the uniform moments [see the vertical axes in Figs.~\ref{fig:MT_curves}(a) and (b)]. 
This suggests that the magnetic field primarily induces staggered moments, which reach the order of $\mu_{\textrm{B}}$ at several Tesla, and the ferromagnetic component follows as a secondary manifestation. 
Comparison of Figs.~\ref{fig:MH_curves}(a) and (b) finds that the staggered moments grow more rapidly than the ferromagnetic moments with increasing magnetic field. 
We note that, at 2 and 3 K, exceptionally large moments are induced by even a low field of 0.3 T, caused by an additional increase of the moment on cooling below $\sim$5 K [Fig.~\ref{fig:MT_curves}(a)]. 
As the magnetization vanishes in the zero-field limit even at these temperatures [Fig.~\ref{fig:MH_curves}(b)], we consider that the staggered moments also vanish at a zero field but appear immediately by applying low fields. 
Later, we will discuss this low-temperature and low-field behavior, taking account of the time scales of the moment fluctuations and NMR observation.

\subsection{$^{1}$H nuclear spin-lattice relaxation}
\begin{figure}
	\includegraphics{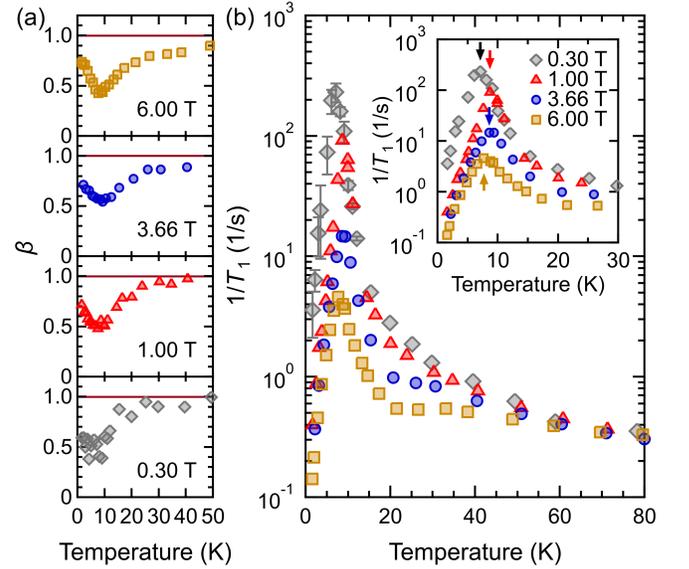}
	\caption{\label{fig:T1} (a) The exponent $\beta$ obtained from the streched exponential fit of the nuclear magnetization recovery. The horizontal lines of $\beta = 1$ correspond to the homogeneous limit. (b) Temperature dependence of the $^{1}$H nuclear spin-lattice relaxation rate, $1/T_{1}$. The enlarged plot below 30 K is shown in the inset. The arrows indicate the peak temperature, $T_{\textrm{peak}}$.}
\end{figure}

Next, we present the nuclear spin-lattice relaxation rate, $1/T_{1}$, which probes the spectral density of spin fluctuations at the NMR frequency. 
The $T_{1}$ was determined from the nuclear-magnetization recovery over time, $M(t)$, after its saturation, namely, $M(0) = 0$. The $M(t)$ was a single exponential function of $t$ at temperatures above 40 K, however, it gradually takes on the non-single exponential feature at lower temperatures. 
(For the relaxation curves, see Appendix~\ref{appendix:relaxation}.)
Thus, we determined $T_{1}$ with fitting the data of $M(t)$ by the stretched exponential function, $1-M(t)/M(\infty) = \exp[-(t/T_{1})^\beta]$, where the exponent, $\beta$, characterizes the degree of inhomogeneity, which approaches 1 in the homogeneous limit. 
In every measured magnetic fields, values of $\beta$ stay in the range of $\beta \leq 0.9$ at high temperatures, however it gradually decreases with temperature below $\sim$30 K, evidencing the evolution of inhomogeneities in spin dynamics at low temperatures as shown in Fig.~\ref{fig:T1}(a). 
In the magnetic fields of 1.00--6.00 T, $\beta$ reaches a minimum value around 7--10 K and, on further cooling, turns to an increase, which seemingly suggests a recovery to the homogeneous relaxation. However, it is presumably not the case but results from the averaging of inhomogeneous $1/T_{1}$ by the spin-spin relaxation (the so-called $T_{2}$ process), which is particularly effective when $1/T_{1}$ is small.

\begin{figure}
	\includegraphics{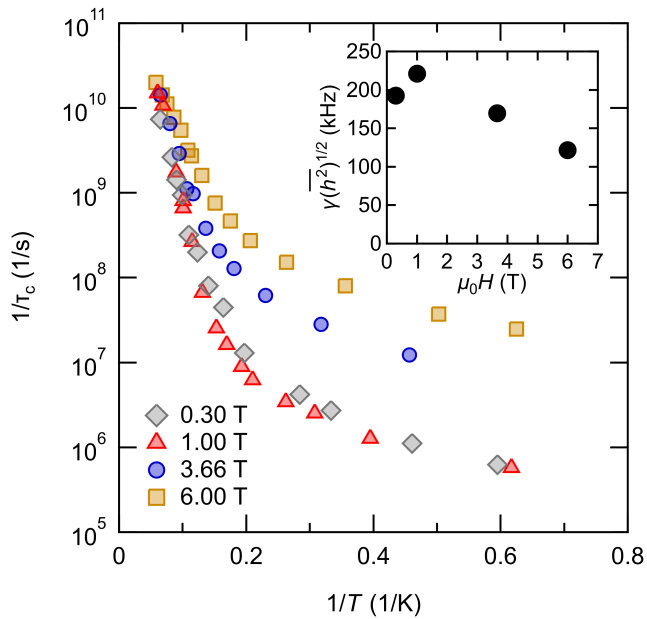}
	\caption{\label{fig:correlation_time} Arrhenius plot of the inverse of correlation time of spin fluctuations, $\tau_{c}(T)$, which was derived from the BPP analyses (see text) of the $1/T_{1}$ data shown in Fig.~\ref{fig:T1}(b). (Inset) Field variation of the amplitude of local field fluctuations, $\gamma \sqrt{\overline{h^{2}}}$.}
\end{figure}

Figure~\ref{fig:T1}(b) shows the temperature dependence of $1/T_{1}$ in the magnetic fields of 0.30, 1.00, 3.66 and 6.00 T. 
Upon cooling from 80 K, $1/T_{1}$ gradually increases and forms a prominent peak at 7--9 K depending on the applied field. 
The $1/T_{1}$ values at low temperatures are strongly suppressed as the magnetic field is increased from 0.30 T to 6.00 T. 
At the first sight, the peak formation of $1/T_{1}$ is reminiscent of a magnetic transition; however, the observed features are distinctive from the manifestations of a magnetic transition as follows. 
First, in the case of magnetic transitions, $1/T_{1}$ forms a peak at a temperature, $T_{c}$, at which spin moments start to emerge, as is observed in $\kappa$-Cu-Cl~\cite{PRL-1995-Miyagawa}, because, on cooling from above $T_{c}$ the spin fluctuations slow down to cause a critical increase in $1/T_{1}$ near $T_{c}$ and upon further cooling below $T_{c}$, where ordered moments are developed, spin-wave excitations responsible for $1/T_{1}$ are depressed. 
In the present results, the moments start to develop at 15, 25, 40 and 60 K in the fields of 0.30, 1.00, 3.66 and 6.00 T, respectively, [see Fig.~\ref{fig:MH_curves}(a)], whereas $1/T_{1}$ forms peaks at much lower temperatures where the moments are sufficiently developed. 
This feature is compatible with our view that the moment generation [Fig.~\ref{fig:MT_curves}(a)] is not due to a spontaneous spin order but is induced by the magnetic field. 
Secondly, the peak value of $1/T_{1}$ in 0.30 T is as large as $\sim$200 1/s, two orders of magnitude beyond the critically enhanced peak value at the AFM transition in $\kappa$-Cu-Cl~\cite{PRL-1995-Miyagawa}. 
Furthermore, a decrease of the $1/T_{1}$ peak value by orders of magnitude with increasing the field to 6.00 T is another extraordinary feature. 
These observations indicate that inhomogeneous, gigantic, and extremely field-susceptible fluctuations free from order prevail in the present spin system.
A possible way to further look into these extraordinary features of $1/T_{1}$, as suggested by Le \textit{et al.}~\cite{PRB-2020-Le}, is the analysis based on the so-called Bloembergen-Purcell-Pound (BPP)-type mechanism, which describes $1/T_{1}$ in terms of the temperature-dependent correlation time of fluctuations, $\tau_{c} (T)$, as~\cite{Slichter,PR-1947-BPP}, 
\begin{equation}
	\frac{1}{T_{1}} \sim \gamma^{2} \overline{h^{2}} \frac{\tau_{c}(T)}{1 + (2\pi f_{0})^{2}\tau_{c}(T)^{2}}%
	\label{eq:BPP}
\end{equation}
with the nuclear gyromagnetic ratio $\gamma$, the mean square of local-field fluctuations, $\overline{h^{2}}$, and the nuclear Larmor frequency $f_{0}$. 
In general, the $\tau_{c}(T)$ increases upon cooling and, when it reaches the value of $(2\pi f_{0})^{-1}$, $1/T_{1}$ forms a peak with the value of $\gamma^{2} \overline{h^{2}} (2\pi f_{0})^{-1}/2$ at a temperature, $T_{\textrm{peak}}$. 
Above $T_{\textrm{peak}}$ where $\tau_{c}(T) \ll (2\pi f_{0})^{-1}$, $1/T_{1} \sim \gamma^{2} \overline{h^{2}} \tau_{c}(T)$, which is $f_{0}$-independent whereas, below $T_{\textrm{peak}}$ where $\tau_{c}(T) \gg (2\pi f_{0})^{-1}$, $1/T_{1} \sim \gamma^{2} \overline{h^{2}} (2\pi f_{0})^{-2}\tau_{c}(T)^{-1}$, which is inversely proportional to $f_{0}^{2}$. 
In the present results, the low-field $1/T_{1}$ data for 0.30 T ($f_{0}$ = 12.8 MHz) and 1.00 T ($f_{0}$ = 42.4 MHz) nearly coincide above $T_{\textrm{peak}}$ and differ by a factor of 10 [nearly corresponding to $(42.4/12.8)^{2}$] below $T_{\textrm{peak}}$ [Fig.~\ref{fig:T1}(b)], supporting the validity of the BPP description. 
When the magnetic field is increased up to 6 T, $1/T_{1}$ is depressed over temperatures across $T_{\textrm{peak}}$, which suggests changes in $\gamma^{2}\overline{h^{2}}$ and(or) $\tau_{c}(T)$ by the high fields. 
Figure~\ref{fig:correlation_time}(b) displays $\gamma\sqrt{\overline{h^{2}}}$ (inset) and $\tau_{c}(T)$ (main panel) derived such that, first, $\gamma\sqrt{\overline{h^{2}}}$ for each magnetic field is evaluated from the peak value through $1/T_{1} = \gamma^{2}\overline{h^{2}}(2\pi f_{0})^{-1}/2$, then, $\tau_{c}(T)$ is calculated from Eq.~\ref{eq:BPP}, using the $\gamma \sqrt{\overline{h^{2}}}$ value for each field. 
The $\gamma \sqrt{\overline{h^{2}}}$ for the low fields of 0.30 and 1.00 T is $\sim$200 kHz, which roughly corresponds to the local field generated by one $\mu_{\textrm{B}}$ on an ET dimer (see the previous subsection). 
This means that the moments fluctuate in all directions. The $\gamma \sqrt{\overline{h^{2}}}$ decreases with increasing field to 3.66 and 6.00 T, which is ascribable to the preferential orientation of the fluctuating spins towards the field direction in the paramagnetic state. 
The $1/\tau_{c}(T)$ is of $\sim$10$^{10}$--10$^{11}$ 1/s at high temperatures, and rapidly decreases with temperature below $\sim$15 K. 
The $1/\tau_{c}(T)$ is field-dependent below $\sim$15 K, where the moments and magnetization are non-linearly induced by magnetic field (see Fig.~\ref{fig:MH_curves}). 
On application of low magnetic fields of, 0.30 and 1.00 T, $1/\tau_{c}(T)$ drops down to below the MHz range at low temperatures. 
This explains the apparent signatures of moments freezing in low fields of 0.30 and 1.00 T at 2 and 3 K [see Fig.~\ref{fig:MH_curves}(a)]; when the NMR observation time scale is faster than the fluctuation speed, NMR sees the instantaneous local-field distribution instead of the motionally averaged one. 
This is further supported by the additional increase in linewidth, $\langle \Delta f^{2}\rangle_{\textrm{m}}^{1/2}$, below $\sim$5 K where $1/\tau_{c}(T)$ is well below the NMR observation frequency in 0.30 and 1.00 T [Fig.~\ref{fig:MT_curves}(a)]. 
Namely, the moments likely keep unfrozen at least down to 2 K, consistent with the magnetization behavior [Fig.~\ref{fig:MH_curves}(b)]. 
At higher fields, 3.66 and 6.00 T, the $1/\tau_{c}(T)$ values are one or two orders of magnitude larger than those at the lower fields, suggesting that the fluctuations become stiffer in the strong magnetic fields, which force the moments to orient to the field direction. 
Intriguingly, $1/\tau_{c}(T)$ does not show the Arrhenius-type behavior but appears to tend towards finite values with no symptom of freezing in every magnetic field, reserving the possibility of a quantum liquid phase.

\section{Discussion}
As indicated by the NMR spectra and magnetization, inhomogeneous local moments grow along with magnetization on cooling and with application of magnetic field. Local moments reach values as large as $\mu_{\textrm{B}}$ whereas magnetization is an order of magnitude smaller, which necessitates staggered-like configurations of the moments. 
Both are nonlinearly induced by magnetic field, not spontaneously emerging in the zero-field limit down to the lowest temperature available. 
A picture derived from these results is that spins form locally AFM-configured clusters that are easily polarized by magnetic field but not ordered or frozen spontaneously. 
Furthermore, as signified by NMR relaxation, the spin dynamics slow down to as low as the submegahertz range, which suggests that a large number of spins sway together. 
The present system with a marginal band profile between half-filled and quarter-filled would be on the verge between a Mott insulator with a spin-frustrated triangular lattice and a charge-ordered insulator in which the charge locations are frustrated between the two sites in each dimer. 
Thus, the spatial profile of charges and therefore spin locations can take a huge number of highly degenerate patterns, possibly leading to inhomogeneous clustering of spins.
As illustrated by Fig.~\ref{fig:Linewidth}, the spatial non-uniformity of the local moments progresses on cooling from 50 K down to 20 K, where the maximum value (represented by $\Delta f_{5\textrm{\%}}$) increases more rapidly than the averaged one ($\Delta f_{50\textrm{\%}}$), suggesting the progressive clustering of spins. 
At the same time, the local moments grow as characterized by a Curie-Weiss (C-W) temperature, $\sim$25 K [the inset of Fig.~\ref{fig:MT_curves}(c)], which is consistent with the first-principles calculations that project an exchange interaction of $\sim10^{2}$ K between spins on ET molecules~\cite{npj-2021-Yamashita} or ET dimers~\cite{PRB-2020-Jacko}. 
Thus, in the temperature range of 20--50 K, the spin clusters gradually take shape with middle-scale staggered spin configurations developed. Remarkably, the uniform magnetization also develops with the similar C-W temperature, $\sim$20 K, on cooling [inset of Fig.~\ref{fig:MT_curves}(d)]. 
The coexistence of \textit{``antiferromagnetic''} staggered moments and \textit{``ferromagnetic''} uniform moments that grow with the similar C-W temperatures is unusual. 
A possible way to solve the apparent contradiction is in terms of the Dzyaloshinskii-Moriya (DM) interaction, which was shown to cause weak ferromagnetism due to spin canting in the AFM state in $\kappa$-Cu-Cl~\cite{PRB-2003-Smith,JPSJ-2018-Ishikawa,PRB-2008-Kagawa} and $\kappa$-Cu-Br~\cite{PRB-2020-Oinuma}. 
However, this mechanism appears not relevant to the present case as follows. 
The spin orbit interaction, which gives the DM interactions, is so small in the ET compounds that the observed canting angle is only 0.3 degrees~\cite{PRB-2003-Smith,JPSJ-2018-Ishikawa,PRB-2008-Kagawa,PRB-2020-Oinuma}. 
In the present system, the ratio of the uniform to staggered magnetizations, $\sim1/10$, corresponds to the canting angle of $\sim$3 degrees, which is too large to be explained by the DM interactions. 
Furthermore, our mean-field calculations of the uniform magnetization of the AFM Heisenberg spins in the presence of the DM interactions, following Ref.~\cite{PRB-2008-Kagawa}, show that the apparent Weiss temperature is negative irrespectively of the magnitude of the DM interactions (except the unrealistic situation that it exceeds the nearest-neighbor exchange interaction), which is adverse to the experimental observation (see Appendix~\ref{appendix:DM} for details).
The viewpoint of the spin cluster formation gives an insight into an alternative and more plausible way to understand the experimental features. 
Spin clusters form as an aggregation of inhomogeneously configured spins, which define an effective spin for each cluster. 
The magnitude of the effective spin should reflect the sizes of the constituent local moments, which would result in the similar Weiss temperatures for the uniform magnetization and local AFM moments. 
For the positive Weiss-temperature, M. Yamashita \textit{et al.}~\cite{npj-2021-Yamashita} have proposed an interesting microscopic mechanism that causes the ferromagnetic interaction between one-dimensional spin fragments. 
The model assumes a stripy charge order hosting one-dimensional finite size spin singlet chains, interrupted by non-ordered defect-like spins whose locations are fluctuating between two ET sites within an ET dimer (called dimer spins). 
Then, the model predicts that the dimer spins mediate the ferromagnetic coupling between the right and left -side adjacent charge-ordered spins, which induce spatially varying staggered moments in the spin singlet chains. 
This model appears to explain the experimental key features such as the formation of inhomogeneously staggered moments and the ferromagnetic magnetization growth in magnetic fields. 
We note, however, that the sizes of the field-induced local moments are as large as the order of $\mu_{\textrm{B}}$ on average, which necessitates very short one-dimensional spin fragments divided by dense dimer spins to be consistent with this model.
Finally, we mention that the overall spatiotemporal features of spin cluster formation, namely, the non-critical slowing down of spin fluctuations concomitant with the significant growth of dynamical correlation length are reminiscent of soft matter~\cite{Jones}. 
The present findings suggest a possible way to share a similar concept between quantum magnetism and soft-matter physics.

\section{Conclusion}
In order to elucidate the microscopic spin states in the dipole liquid candidate, $\kappa$-(ET)$_{2}$Hg(SCN)$_{2}$Br, we conducted $^{1}$H NMR study in magnetic fields varied over a 20-fold range. 
The field and temperature variations of NMR spectra indicate that staggered moments are inhomogeneously and nonlinearly induced by the magnetic field, reaching the order of $\mu_{\textrm{B}}$ per ET dimer in magnetic field of a few Tesla at low temperatures. 
The NMR $1/T_{1}$ exhibits temperature variations with pronounced peaks at 7--9 K ascribable to the non-critical slowing down of spin fluctuations instead of the critical enhancement associated with the spontaneous magnetic ordering. 
The analyses of the NMR spectra and $1/T_{1}$, in conjunction with the magnetization behavior, leads to the following picture of the spin state in $\kappa$-(ET)$_{2}$Hg(SCN)$_{2}$Br: The spins with short-ranged AF correlation at high temperatures develop their moments heterogeneously upon cooling down to $\sim$20 K so that inhomogeneous spin clusters are shaped and grow. 
Below $\sim$20 K, the dynamics of the spin clusters gradually slows down to frequencies below MHz range without any signature of spin freezing down to 2 K. 
The observed inhomogeneity, short or middle-range correlations and slow dynamics are indicative of the soft-matter nature of the present spin system. 
Intriguingly, the slowing-down of the spin fluctuations is not of activation type but tends to cease at low temperatures with the correlation time saturating to finite values. 
This slowly fluctuating spin clusters maintaining liquidity even at low temperatures can be a novel type of spin organization free from ordering or freezing that emerges in a system with charge and spin frustrations.

\begin{acknowledgments}
We thank C. Hotta, M. Yamashita, S. Uji, S. Sugiura, and S. Dekura for fruitful discussions. 
A part of the experiments was performed using facilities of the Cryogenic Research Center, University of Tokyo. 
This work was supported in part by JSPS Grants-in-Aid for Scientific Research (Grants Nos. JP18H05225, 20K20894, 20KK0060 and 21K18144). 
N.D. is grateful for the support of the Visiting Researcher's Program of the Institute for Solid State Physics, University of Tokyo, and NSF award DMR-2004074. 
The work in Chernogolovka was carried out within the state assignment of the Ministry of Science and Higher Education of Russian Federation (number AAAA-A19-119092390079-8).
\end{acknowledgments}

\appendix

\section{Numerical calculation for the second moment of $^{1}$H NMR spectra}\label{appendix:SPCsimulation}
As mentioned in the main text, the second moment of the $^{1}$H NMR spectra, $\langle \Delta f^{2}\rangle$, consists of the temperature-insensitive $^{1}$H nuclear dipolar contribution, $\langle \Delta f^{2}\rangle_{\textrm{n}}$, and the dipolar contribution from the moments of electron spins, $\langle \Delta f^{2}\rangle_{\textrm{m}}$. 
To estimate the average size of the local moments from the experimental values of $\langle \Delta f^{2}\rangle_{\textrm{m}}$, we calculated the $^{1}$H NMR spectra, which is to be related to $\langle \Delta f^{2}\rangle_{\textrm{m}}$, using the reported crystal structure of $\kappa$-Hg-Br~\cite{BRAS-1992-Konovalikhin} (and the isostructural sister compound $\kappa$-Hg-Cl~\cite{PRB-2014-Drichko}) and considering the minimal Hamiltonian describing the interactions between the pair of the $^{1}$H nuclear spins with $I_{ji}$ ($i$ = 1, 2) bonded to the carbon $j$ and the electronic spins in the following:
\begin{widetext}
\begin{equation}
	\mathcal{H}^{j} = \sum_{i}(-\gamma\hbar{\delta H}_{ji}^{z}I_{ji}^{z}) + \frac{\gamma^{2}\hbar^{2}}{r_{i}^{3}} \left[ I_{j1}^{z}I_{j2}^{z} - \frac{1}{4}(I_{j1}^{+}I_{j2}^{-} + I_{j1}^{-}I_{j2}^{+})\right] (1 - 3\cos^{2} \phi_{j}),%
	\label{eq:dipoleHamiltonian}
\end{equation}
\end{widetext}
where $g$ and $\hbar$ are the gyromagnetic ration of $^{1}$H nuclei and the reduced Planck constant. 
The first term is the hyperfine interactions between each $^{1}$H nuclear spin and all the electronic moments; because the magnitude of the hyperfine field, $|\bm{\delta H_{ji}}|$, is much smaller than that of the applied field, the interaction is well approximated by its $z$ component. 
The second term comes from the dipole interactions between a pair of $^{1}$H nuclear spins under consideration. 
The distance between the $^{1}$H nuclei and the angle between the vector connecting the paired $^{1}$H nuclei and an applied magnetic field are denoted as $r_{j}$ and $\phi_{j}$, respectively. 
In the present case, the hyperfine field, $\bm{\delta H_{ji}}$, mainly comes from the electronic dipolar fields, $\bm{H_{\textrm{dip},ji}}$; so, we approximated it as the sum of the external field-parallel component of the point-dipolar fields from the all the off-site atom positions within a sphere of $\sim$100 {\AA} radius:
\begin{equation}
	\bm{H_{{\textrm{dip}},ji}} = \sum_{k} \sigma_{k} \left\{ 3 \frac{(\bm{\mu_{k}}\cdot\bm{r_{ji,k}})}{r_{ji,k}^{5}} - \frac{\bm{\mu_{k}}}{r_{ji,k}^{3}}\right\},\nonumber
\end{equation}
where $\bm{r_{ji,k}}$ is the vector from the atomic site ($k$) to the proton site ($ji$) and $r_{ji,k} = |\bm{r_{ji,k}}|$. 
The prefactor $\sigma_{k}$ is so-called the Mulliken charge, the population density of the highest occupied molecular orbital (HOMO), at atom k calculated by the extended H\"{u}ckel method~\cite{BullChem-1984-Mori}, and $\mu_{k}$ is the total moment of the ET molecule that atom $k$ belongs to. 
In the present simulation, we ignored the isotropic Fermi contact from on-site electronic moment or the transferred hyperfine couplings via $\sigma$ bond to the neighboring carbon because it is, in general, negligibly small in the ET compounds due to the small HOMO densities in the ethylene group that reside at the edges of ET molecules. 
Indeed, for a radical ET$^{+}$, the hyperfine coupling constant at the $^{1}$H site is less than $2\times10^{1}$ Oe/$\mu_{\textrm{B}}$~\cite{1986-Cavara}. 
We also ignored the moment-independent chemical shift. 
Note that the present simulations do not consider the demagnetization effect because the uniform magnetization is much smaller than expected from the fully polarized moments and the main contribution to the local fields is from the staggered-like moments. 
(For reference, the demagnetization factors estimated by the spheroidal approximation of the sample dimension, $1 \times 0.4 \times 0.4$ mm$^{3}$, are $\sim$0.43 (0.14) for applying the magnetic field along the short (long) axis of the sample; the field direction in the present study is along one of the short axes.)
From Eq.~\ref{eq:dipoleHamiltonian}, we expect the quartet spectrum from each pair of protons bonded to the same carbon:
\begin{subequations}
\label{eq:quartet}
\begin{eqnarray}
	\Delta E_{1\leftrightarrow 2} = D_{j+} - \sqrt{D_{j-}^{2} + J_{j}^{2}} - 2J_{j},
	\\
	\Delta E_{3\leftrightarrow 4} = D_{j+} - \sqrt{D_{j-}^{2} + J_{j}^{2}} + 2J_{j},
	\\
	\Delta E_{1\leftrightarrow 3} = D_{j+} + \sqrt{D_{j-}^{2} + J_{j}^{2}} - 2J_{j},
	\\
	\Delta E_{2\leftrightarrow 4} = D_{j+} + \sqrt{D_{j-}^{2} + J_{j}^{2}} + 2J_{j}, 
\end{eqnarray}
\end{subequations}
with the intensities
\begin{subequations}
\label{eq:Intensity}
\begin{eqnarray}
	I_{1\leftrightarrow 2} = 1 + \sin 2\alpha, 
	\\
	I_{3\leftrightarrow 4} = 1 - \sin 2\alpha, 
	\\
	I_{1\leftrightarrow 3} = 1 - \sin 2\alpha,
	\\
	I_{2\leftrightarrow 4} = 1 + \sin 2\alpha,
\end{eqnarray}
\end{subequations}
where
	\begin{eqnarray}
		D_{ji\pm} = \frac{1}{2} \gamma \hbar \left(\delta H_{j1}^{z} \pm \delta H_{j2}^{z} \right),\nonumber
		\\
		J_{j} = \frac{\gamma^{2} \hbar^{2}}{4r_{j}^{3}} \left(1 - 3\cos^{2} \theta_{j} \right),\nonumber
		\\
		\tan 2\alpha_{j} = \left|\frac{J_{j}}{D_{j-}} \right|.\nonumber
	\end{eqnarray}
The total spectrum for an ET molecule consists of the 16 lines from the inequivalent four pairs of protons. 
We assumed the Gaussian shape $g(f)$ for each of the spectral lines $n$ ($n$ = 1, 2, ..., 16) as the form of $I_{n} g(f - f_{n0})$, where $f_{n0}$ and $I_{n}$ are the central position and its intensity of line $n$ determined by Eqs.~\ref{eq:quartet} and~\ref{eq:Intensity}, respectively. 
We performed these calculations for magnetically inequivalent ET molecules and constructed the numerical spectrum by superposing all the lines. 
Given the pattern of the spin configuration, the simulated $\langle \Delta f^{2}\rangle_{\textrm{m}}^{1/2}$, determined by the second moment of the spectrum subtracted by that of the spectrum for zero local moments, is proportional to the size of the local spin moments, $m_{\textrm{local}}$; $\langle \Delta f^{2}\rangle_{\textrm{m}}^{1/2} = am_{\textrm{local}}$ with the coefficient a depending on the moment configuration.

\begin{figure*}
	\includegraphics{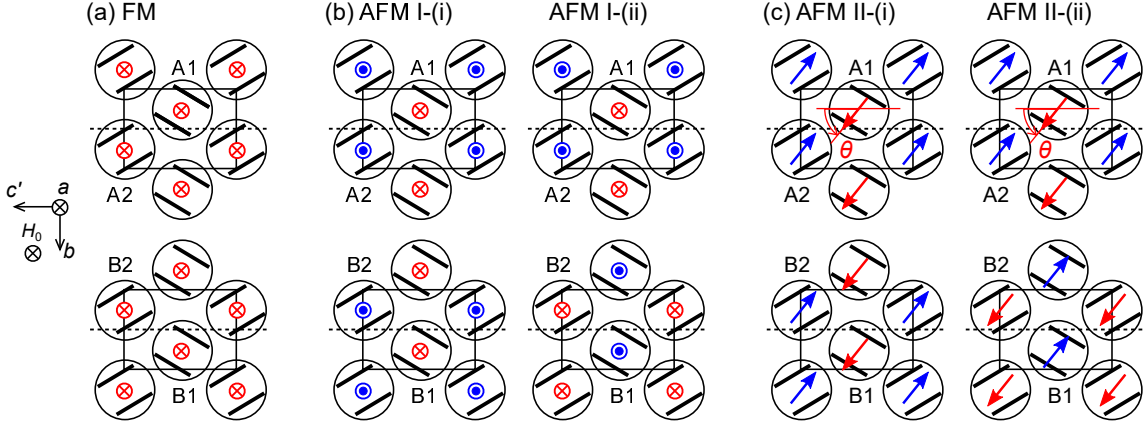}
	\caption{\label{afig:mom_config} The moment configurations considered in the spectral simulations in the present work. The solid line rectangles and dashed lines represent the unit cell and the glide $ac^{\prime}$ plane, respectively.}
\end{figure*}

Figure~\ref{afig:mom_config} and Table~\ref{tab:simulation} show the moment configurations considered in the present work and the values of $a$ for each configuration. 

\begin{table}
	\caption{\label{tab:simulation} The coefficient a in unit of kHz/$\mu_{\textrm{B}}$ per ET dimer calculated for the moment configurations considered in the present work shown in Fig.~\ref{afig:mom_config}.}
	\begin{ruledtabular}
	\begin{tabular}{ccccc}
		& & & $\kappa$-Hg-Br~\cite{BRAS-1992-Konovalikhin} & $\kappa$-Hg-Cl~\cite{PRB-2014-Drichko}
		\\
		\hline
		FM & & & 181 & 179\\
		\hline
		AFM I & I-(i) & & 195 & 170\\ \cline{2-5}
		& I-(ii) & & 156 & 140\\ \hline
		AFM II & II-(i) & $\theta = 0^{\circ}$ & 221 & 245\\ \cline{3-5}
		& & $\theta = 90^{\circ}$ & 106 & 102\\ \cline{2-5}
		& II-(ii) & $\theta = 0^{\circ}$ & 238 & 261\\ \cline{3-5}
		& & $\theta = 90^{\circ}$ & 107 & 106
	\end{tabular}
	\end{ruledtabular}
\end{table}
As representatives of the possible staggered moment configurations, we considered two simple colinear configurations where local moments point parallel [``AFM-I'' in Fig.~\ref{afig:mom_config}(b)] and perpendicular [``AFM-II'' in Fig.~\ref{afig:mom_config}(c)] to the applied magnetic field, respectively. 
We considered both the ferromagnetic (i) and antiferromagnetic (ii) interlayer configurations, because the sign of the interlayer interaction depending on the compound~\cite{PRB-2020-Oinuma} is not known in the present compound. 
For the configuration ``AFM-I'', we approximated all the local moments to have the same $m_{\textrm{local}}$ for simplicity. 
For the configuration ``AFM-II'' in Fig.~\ref{afig:mom_config}(c), we put the in-plane sublattice moments on the two inequivalent dimers in a layer, $m_{1}$ and $m_{2}$ as
\begin{subequations}
\label{eq:sublattice_mom}
\begin{eqnarray}
	m_{1} &=& m_{\textrm{local}} (\bm{\hat{e_{b}}}\sin \theta + \bm{\hat{e_{c}^{\prime}}}),\nonumber\\
	m_{2} &=& m_{\textrm{local}} \{\bm{\hat{e_{b}}}\sin (\theta + \pi) + \bm{\hat{e_{c}^{\prime}}}\cos (\theta + \pi)\},\nonumber
\end{eqnarray}
\end{subequations}
where $\bm{\hat{e_{b}}}$ is the unit vector of $b$ axis and $\bm{\hat{e_{c}^{\prime}}}$ is the unit vector of $c$ axis projected to the plane perpendicular to a axis; $\bm{\hat{e_{c}}} = \bm{\hat{e_{a}}}\cos \beta + \bm{\hat{e_{c}^{\prime}}}\sin \beta$ with the angle between $a$ and $c$ axes, $\beta = 89.71^{\circ}$ for $\kappa$-Hg-Br~\cite{BRAS-1992-Konovalikhin}. 
In case of the intradimer spin-density imbalance, the NMR spectra were calculated in the same way as described above except with the spin density disproportionated in the dimer (see Fig.~\ref{fig:SPCsimulation} in the main text).

\section{Recovery curves of $^{1}$H nuclear magnetization}\label{appendix:relaxation}
\begin{figure*}
	\includegraphics{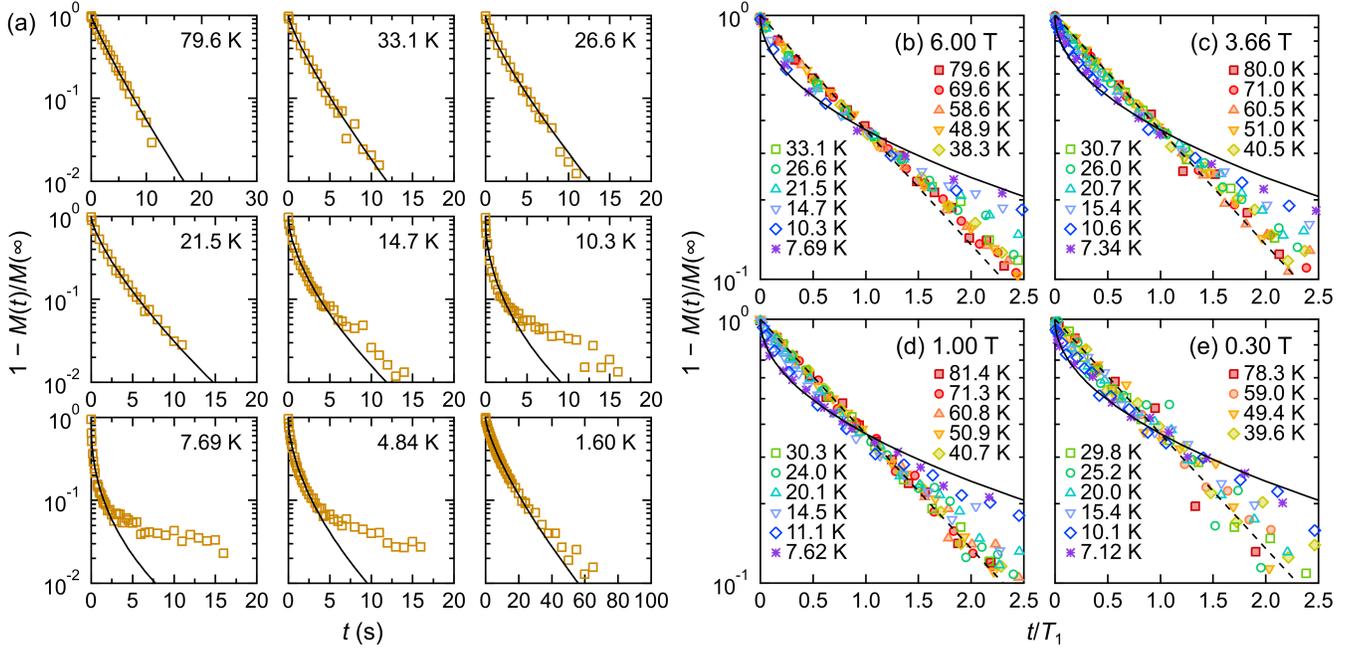}
	\caption{\label{afig:relaxation_all} (a) Recovery curves of $^{1}$H nuclear magnetization at various temperatures in the magnetic field of 6.00 T. The solid curves are the fits of the stretched exponential function to the entire recovery curves. (b--e) Recovery curves of the nuclear magnetization plotted against $t/T_{1}$ above 7 K in the fields of (b) 6.00, (c) 3.66, (d) 1.00 and (e) 0.30 T. The broken lines and solid curves represent the functions of $\exp(-t/T_{1})$ and $\exp[-(t/T_{1})^{0.5}]$, respectively.}
\end{figure*}

Figure~\ref{afig:relaxation_all} shows the recovery curves of $^{1}$H nuclear magnetization measured at various temperatures below $T_{\textrm{MI}}$ in the magnetic field of 6.00 T. 
The recovery curves show bending at low temperatures, indicating the developing inhomogeneity in the nuclear spin lattice relaxation rate $1/T_{1}$, most visible around 7--10 K. 
To compare the bending nature of the recovery curves at different temperatures, we plot the relaxation curve of $1 - M(t)/M(\infty)$ versus $t/T_{1}$ at 6 T for different temperatures in Fig.~\ref{afig:relaxation_all}(b), where $t/T_{1}$ is a time normalized to $T_{1}$ determined by the fit of the data of $1 - M(t)/M(\infty)$ by the stretched exponential function, $\exp[-(t/T_{1})^{\beta}]$, with the exponent $\beta$ characterizing the degree of inhomogeneity [see also Fig.~\ref{fig:T1}(a)]. 
At high temperatures, the relaxation curve is close to the single exponential function, $1 - M(t)/M(\infty) = \exp(t/T_{1})$, as expected for a homogeneous relaxation. 
On cooling, the recovery curves gradually deviate from the single exponential function and approach the form, $1 - M(t)/M(\infty) = \exp[-(t/T_{1})^{0.5}]$. 
Figures~\ref{afig:relaxation_all}(c), (d) and (e) show the results for the applied magnetic fields of 3.66, 1.00 and 0.30 T, respectively. 
In each magnetic field, the temperature variation of the recovery curve are similar to those in 6.00 T, as seen in that of $\beta$ in Fig.~\ref{fig:T1}(a).

\section{Field-induced moments in the presence of the DM interaction}\label{appendix:DM}
\begin{figure*}
	\includegraphics{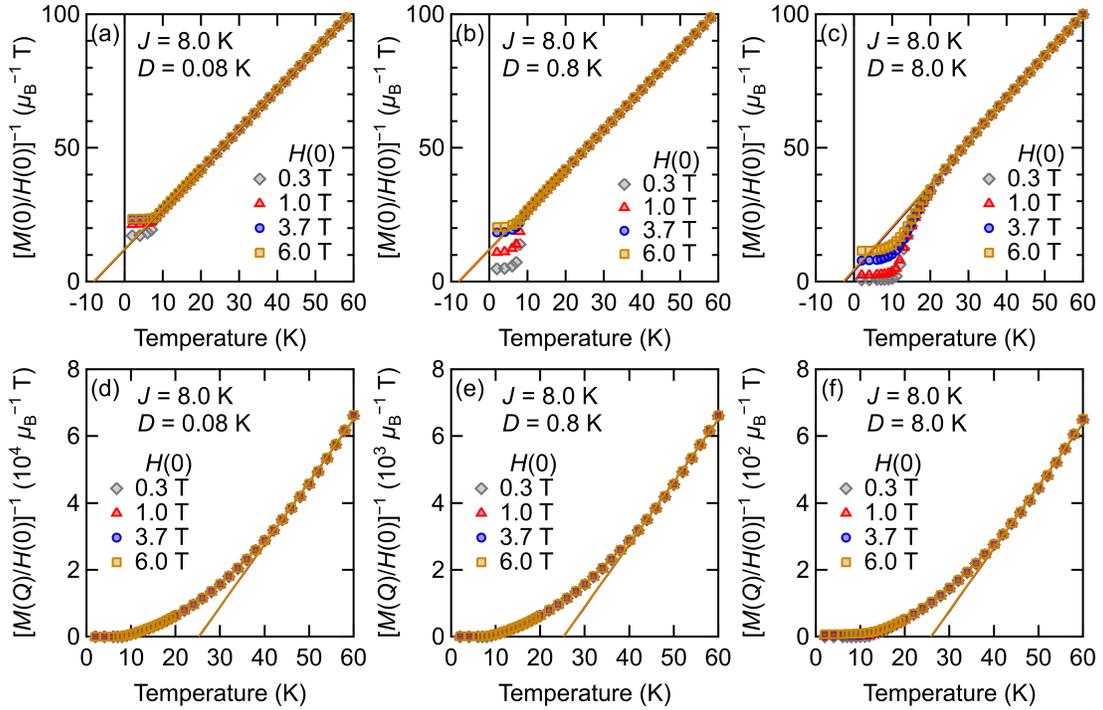}
	\caption{\label{afig:DM_simulation} The plots of the inverse of the calculated uniform and staggered magnetization divided by the external magnetic field, $[M(0)/H(0)]^{-1}$ (a--c) and $[M(Q)/H(0)]^{-1}$ (d--f) versus temperature in various external magnetic fields in the presence of the DM interaction, respectively. The solid lines are the Curie-Weiss fits in the temperature range between 40 and 60 K.}
\end{figure*}

To verify if the DM interaction explains the field-induced magnetism in $\kappa$-Hg-Br, we simulated the uniform and staggered moments in the presence of the DM interaction under magnetic fields within the molecular-field theory, following Ref.~\cite{PRB-2008-Kagawa}. 
We considered the $S = 1/2$ antiferromagnetic (AFM) Heisenberg model on a two-dimensional square lattice in the $xy$-plane described by
\begin{equation}
	\mathcal{H} = J \sum_{\langle i,j\rangle} \bm{S_{i}} \cdot \bm{S_{j}} + g\mu_{\textrm{B}}\sum_{i} \bm{S_{i}} \cdot \bm{H_{i}} + \sum_{\langle i,j\rangle} \bm{D_{ij}} \cdot \left(\bm{S_{i}} \times \bm{S_{j}}\right),\nonumber%
	\label{eq:DM_Hamiltonian}
\end{equation}
where $\bm{S_{i}}$ and $\bm{H_{i}}$ are the spin operator and the external magnetic field at site $i$, respectively. 
The nearest-neighbor exchange interaction and DM vector are denoted as $J$ and $\bm{D_{ij}}$. 
If $\bm{H_{i}} = \bm{D_{ij}} = 0$, the system exhibits the N\'{e}el order with a wave-number vector of $(\pi,\pi)$ to form the two sublattices, $+$ and $-$, with the same magnitude of the moments. 
For simplicity, we assume that $\bm{D_{ij}}$ is parallel to the $z$-axis. 
The uniform external magnetic field $\bm{H}(\bm{0})$ is set to be parallel to the $y$-axis.
The effective fields at the sublattices $+$ and $-$, $\bm{H_{+}^{\textrm{eff}}}$ and $\bm{H_{-}^{\textrm{eff}}}$, are written as
\begin{subequations}
	\label{eq:H_eff}
	\begin{eqnarray}
		\bm{H_{+}^{\textrm{eff}}} = -\frac{ZJ}{(g\mu_{\textrm{B}})^{2}} \langle \bm{M_{-}}\rangle - \frac{Z}{(g\mu_{\textrm{B}})^{2}} \langle \bm{M_{-}} \rangle \times \bm{D} + \bm{H_{i}},\nonumber\\
		\bm{H_{-}^{\textrm{eff}}} = -\frac{ZJ}{(g\mu_{\textrm{B}})^{2}} \langle \bm{M_{+}}\rangle - \frac{Z}{(g\mu_{\textrm{B}})^{2}} \langle \bm{M_{+}} \rangle \times \bm{D} + \bm{H_{i}},\nonumber
	\end{eqnarray}
\end{subequations}
with the number of nearest-neighboring sites $Z = 4$, the DM vector $D = (0, 0, -D)$, the external magnetic field $\bm{H_{i}} = (0, H(0), 0)$. 
Here $\langle \bm{M_{+}} \rangle$ ($\langle \bm{M_{-}} \rangle$) denotes the averaged moment of the sublattice $+$ ($-$). 
The molecular-field solutions were obtained by assuming  $\langle \bm{M_{+}} \rangle || \bm{H_{+}^{\textrm{eff}}}$ and $\langle \bm{M_{-}} \rangle || \bm{H_{+}^{\textrm{eff}}}$. 
Due to the presence of non-zero Zeeman and DM interactions ,$\langle \bm{M_{+}} \rangle$ and $\langle \bm{M_{-}} \rangle$ pointing in the $xy$-plane are not mutually parallel but symmetric across the $y$-axis. 
Below, we discuss the solutions using the basis of $M(\bm{0}) = \left|\langle \bm{M_{+}} \rangle + \langle \bm{M_{-}} \rangle\right|/2$ and $M(\bm{Q}) = \left|\langle \bm{M_{+}} \rangle - \langle \bm{M_{-}} \rangle\right|/2$, which represent uniform and staggered moments along the $y$- and $x$- axes, respectively. 
Throughout the simulations, the exchange interaction was set to $J$ = 8 K so that it coincides the peak temperature in $1/T_{1}$. 
When $H(0) = 0$, the non-zero $M(\bm{Q})$ emerges below $T_{\textrm{N}} \sim (J^{2} + D^{2})^{1/2}$ and $M(\bm{0})$ also becomes finite below $T_{\textrm{N}}$ only when $D \neq 0$. 
In the non-zero $H(0)$, $M(0)$ seemingly follows the Curie-Weiss' law in the nominal paramagnetic state at high temperatures for $D \leq J$ [Figs.~\ref{afig:DM_simulation}(a)--(c)]. 
The apparent Weiss temperatures of the $M(0)$ behavior in 40--60 K under 0.30 T for $D$ = 0.08, 0.8, and 8.0 K are $\Theta_{\textrm{W}}$ = $-8.0$, $-8.0$, and $-2.4$ K, respectively. 
For a fixed $D$ satisfying $D \leq J$, the inverse of the uniform magnetic susceptibility, $\left[M(0)/H(0)\right]^{-1}$ is less dependent on $H(0)$ at high temperatures above $\sim$20 K and the sign of $\Theta_{\textrm{W}}$ is always negative for $H(0)$ up to 6.0 T. 
These numerical results qualitatively differ from the experimental behavior of the uniform magnetic susceptibility showing a positive $\Theta_{\textrm{W}}$ value of $\sim$20 K [Fig.~\ref{fig:MT_curves}(d)], whereas both the calculated staggered moment, $M(\bm{Q})$, and the experimental local moment evaluated from $\langle \Delta f^{2}\rangle_{\textrm{m}}^{1/2}$ have negative $\Theta_{\textrm{W}}$ temperatures [Figs.~\ref{afig:DM_simulation}(d)--(f) and Fig.~\ref{fig:MT_curves}(c)]. 
Thus, the observed behaviors of the uniform and staggered magnetizations cannot be coherently explained by the simple model including antiferromagnetic exchange interaction and the DM interaction.

\providecommand{\noopsort}[1]{}\providecommand{\singleletter}[1]{#1}%

\end{document}